**Influence of quantum-mechanical boundary roughness resistance on copper nanolines.**

T. E. Huber, P. Trottman, and J. B. Halpern

Howard University, Washington, DC 20059

Copper nanolines fabricated by the damascene process are commonly used as interconnects in advanced electronic devices. The copper resistivity increases above the bulk value because of confinement, in detriment of performance. Recent electronic transport measurements clearly exhibit this effect for nanolines whose widths range between 80 and 500 nm at low and room temperatures. An interpretation in terms of finite size effects that consider semiclassical models for electron-surface and grain-boundary scattering was presented, but the fits do not capture the strong linewidth dependence in the data below 100 nm. The present letter explains how the excess resistivity arises from a quantum mechanical surface roughness effect that begins to contribute strongly in lines narrower than the mean free path. This type of roughness scattering would contribute a 40% increase in resistivity for future 20 nm-interconnects.

As dimensions decrease, the electronic properties of materials reach fundamental limits. A key issue in the continuing evolution of microelectronics is the resistivity increase in Cu lines used for interconnects due to size effects.[1] These copper lines have rectangular cross-sections with the thinner side, the linewidth $w$, measuring tens of nanometers. Their resistivity is relatively low because the electron density of copper is very high. To conduction electrons in a thin wire, the external surfaces play the same role as impurities, resulting in a finite-size effect, a residual resistivity that increases inversely with the linewidth. Plombon, Andideh, Dubin and Maiz[2] presented experimental studies of copper line resistivities for $w$ ranging between 80 nm and 500 nm at room and cryogenic temperatures. Fig.1 (a) presents the data as a log-log plot, showing the expected $w^{-1}$ dependence for large $w$ and evidence of a stronger, $w^{-2}$, dependence for the narrower wires which is more obvious at low temperatures. As commented on by Plombon *et al*, even complex semiclassical (SC) models including finite-size effects and grain scattering fail to naturally capture the excursion of the resistivities from an inverse linear dependence for the narrowest lines in their study.

Quantum mechanical size effects are not generally taken into account for copper nanolines because the Fermi wavelength, $\lambda_F$, is much shorter than the linewidth however, a similar (*thickness*)$^{-2}$ behaviour been observed experimentally in a number of metallic films[3] and has been interpreted in terms of quantum-mechanical roughness scattering (QMRS).[4-6] QMRS is governed by an interaction potential that is derived from the confinement energy that scales as $w^{-2}$. In the cases where strong QMRS is expected, for example for semimetallic Bi nanowires given their long $\lambda_F$ of around 50 nm, the evidence is substantial.[7,8] The present letter considers the merits of quantum mechanical roughness as the source of the observed anomalous resistivity increase in copper nanolines.

When the linewidth or film thickness is less than the mean free path, *mfp*, charge carrier scattering at the nanostructure boundary becomes comparable to intrinsic scattering. Thus, the electrical resistivity of the material in a nanostructure will increase from its bulk value. Fuchs,[9] Dingle,[10] and Sondheimer (FDS)[11] presented a very successful theory of finite size effects based on the Boltzmann equation and a model of scattering at the boundaries. Their model is underpinned by the semiclassical concept of specularity, *p*, the probability of "reflective" scattering that changes the sign of the momentum perpendicular to the boundary, leaving the momentum parallel to the boundary unchanged. Mayadas and Shatzkes[12] calculated the additional resistivity increase associated with scattering from the grain boundaries. They calculated the probability of an electron reflecting from a grain boundary, *R*, which arises when the standard deviation $\xi$ of the grain location from the perfect periodic case is long in comparison with the Fermi wavelength $\lambda_F$. Rossnagel and Kuan[13] (RK) additionally took into account surface roughness, arriving at an expression that considers both FDS's electron-surface scattering as well as semiclassical surface roughness scattering in the first term of Eq. 1 below and grain scattering in the second term:

$$\rho_{RK}/\rho_0 \approx \{1 + 0.375(1-p)S \times mfp/w + 1.5[R/(1-R)] \times mfp/g\} \quad (1)$$

$\rho_{RK}$ and $\rho_0$ are the nanoline and bulk value of the conductivity, respectively. *S* is the roughness factor, an empirical constant that is assumed to be unity and *g* is the grain size. If *g* ~*w*, then $\rho_{RK}$ scales naturally as $w^{-1}$. Fig.1 shows Plombon *et al.*'s fit to their data using Eq. 1 where *g* was characterized via scanning electron microscopy and found to be roughly equal to $w$. The electron-surface scattering was assumed to be perfectly diffuse ($p = 0$), the grain boundary reflection coefficient, $R = 0.25$, and the *mfp*s were 39 nm and 300 nm at room temperature and 20 K, respectively. The choice of parameters is not entirely supported by the ancillary experimental results that they presented; still these choices enable them to successfully model the resistivity increase in $\rho_{RK}$ except for the narrowest lines. To get reasonable fits in those cases, they had to postulate that $g$ decreased by as much as 40% from the value for wider lines, an *ad hoc* assumption. We claim that the reason for the poor fit is the absence of a QMRS term, proportional to $w^{-2}$, in their analysis.

Regarding QMRS, theoretical studies of surface-induced scattering[4-6] in thin films indicate that semiclassical approaches fail and a quantum mechanical treatment is required in fine structures as the electron motion is coherent over distances of the order of the *mfp*. Indeed, in copper at low temperature, *mfp* >> *w*, but the case is more doubtful at room temperatures. QMRS is illustrated in Fig. 2. The theoretical approach of Trivedi and Ashcroft[4] handles transport in a rough wire in two parts, each corresponding to a different scale. The effect of roughness at the scale of the *mfp* or less is included by evaluating the conductivity of a segment *i* having small variations in its width *W(i)* about an average value *w*. The large scale fluctuations in the width are subsequently treated semiclassically suggesting that finite size effects and QMRS exist independently. Since the electron wavefunction is coherent over the segment, surface roughness is considered quantum mechanically as an effective interaction potential derived from the boundary condition of the exact Hamiltonian. Fig. 2.b shows a schematic of the band structure of a film representing the nanoline. Because of quantum confinement the levels are separated by $\varepsilon = 2\pi E_F/k_F w$ where $E_F$ is the Fermi energy, 7 eV, and $k_F$ the Fermi wavelength. $k_F = 1.36 \times 10^8$ cm$^{-1}$ ($\lambda_F = 0.46$ nm) as deduced from the electron density of bulk copper. The solid and dashed horizontal lines represent levels of the subband structure of the metal in a film that are filled and unfilled, respectively. The gray band represents thermal fluctuations. The arrow in the inset represents the interband transitions within a band around the Fermi level that is associated with disorder within the coherence length and increase the resistivity:

$$\rho_{QMRS} \approx \frac{2\hbar}{e^2 k_F} (\delta w/w)^2 \qquad (2)$$

For copper, $\frac{2\hbar}{e^2 k_F} = 2380$ μΩ-cm, which is temperature independent. Considering that the correlation length $\zeta$ is 0, this amount to a flat spectrum of spatial periodicities for surface fluctuations. Eq. 2 introduces one adjustable parameter, $\delta w$. Our model considers the combined result of RK finite size effects (Eq. 1) and QMRS (Eq. 2) using the Matthiessen rule $\rho = \rho_{RK} + \rho_{QMR}$. Our best fit, shown in Fig 1 (b), is obtained for $\delta w = 0.8$ nm and $A(T) = (0.375(1-p) + 1.5R/(1-R))mfp$ yielding 31 nm and 286 nm for 300 K and 20 K, respectively. Since $A$ is found, within experimental error, to be proportional to the *mfp*, we find that any combination of $p$ and $R$ that satisfies $(0.375(1-p) + 1.5R/(1-R)) = 0.7$ fits and fits equally well. Combinations of $p$ and $R$ that fit are shown in Fig. 1 (c). For example, a possible solution is to assume $p = 0.1$ (increased slightly from the 0.0 in Plombon *et al*) so that the grain boundary reflection coefficient $R$ is correspondingly reduced from 0.25 to 0.18, leaving all the other parameters the same as in the baseline, Plombon *et al* fit. ($S = 1$ and $g = w$). Our model, that assumes that the samples are fabricated through a process that maintains uniform $\delta w$ as $w$ decreases, fits the entire range of linewidths at the two temperatures without *ad hoc* parametrization.

Next we discuss the RK-QMRS interplay. In our fit, *p* and *R* are not very different than those required to fit the data using Eq. 1 alone. The resistance excess $\rho_{RK}/\rho_0$, at room temperature, is 1.56 in Plombon *et al* and 1.47 in ours. This means that the scattering inherent to Eq. 1 is not decreased, at least not significantly, to compensate for the contribution coming from Eq. 2. In other words, the fitted results support the idea that RK and QMRS scattering exist independently. It is even possible, as follows, to argue that the characteristics of the roughness that contribute to RK and QMRS are different. The phenomenological picture that accompanies Eq. 1 with *S* =1 is that only those electrons contribute to the current whose momentum $\vec{k}_F$ is oriented nearly parallel to the wire axis. Therefore, for the RK scattering, within a single *mfp* one only need consider collisions of the electrons, whose wavelength is $\lambda_F$, with locations on the surface where they can be diffracted away from a trajectory parallel to the wire axis. Consequently the relevant roughness has spatial periodicities of ~$\lambda_F$ times the ratio of the *mfp* to the linewidth which is of the order of $\lambda_F$. On the other hand, the fluctuations of linewidth that can potentially give rise to QMRS are those with spatial periodicity of up to the order of the *mfp*, which is at least two orders of magnitude larger than $\lambda_F$. We have so far considered a first order approximation. In the very narrow lines, when *w* << *mfp*, quantum-mechanical interband transitions, as a second order perturbation, admix the electron subband states. Then, one wonders about quantum mechanical scattering, if it affects, and if it could be used to manipulate, the semiclassical effects. This is very important for the application of nanolines to interconnects.

The most interesting case is the admixing of two wavefunctions of states that are consecutive in the energy spectrum, whose energy difference is $\varepsilon$ (see Fig 2). The superposition of these wavefunctions, the real wavefunctions, have components with a long spatial period, $2m/\hbar\sqrt{\varepsilon} \sim 10^2 \lambda_F$. These components would be scattered only by correspondingly long-period roughness. Therefore, in second order, interband transitions caused by QMRS take away strength from RK.

We assumed that roughness was uncorrelated. Actually, the role of the correlation function and the correlation length $\zeta$ of roughness in scattering has been studied in thin films in cases where $\zeta > 0$ also. It was concluded that the $\rho$ associated with QMRS depends on the choice of correlation function and $\zeta$, both in the case $\zeta > \lambda_F$ and when[5] $\zeta < \lambda_F$ and that the $w^{-2}$ dependence is rather robust to the characteristics of roughness. Further, the same dependence is obtained from general arguments based on electron diffraction.[6] Clearly, studies of electronic transport for Cu nanolines with different roughness are highly desirable. They would be similar to those of Boukai et al and Hochbaum et al[14] for phonon scattering in semiconductor nanowires. Comprehensive electronic transport experimental work (such as magneto-resistance and thermopower) may be required because there are many parameters that need to be determined to understand interconnect resistance both in the semiclassical and quantum limits.

On the theoretical side, motivated by the resistivity increase in copper nanolines, Timoshevskii et al[15] presented an *ab initio* study of the effects associated with quantum mechanical transport in the presence of roughness. Unfortunately, their predictions were not compared with the nanolines resistivity data and the same applies to Feldman et al.[16]

The finite size effect in copper lines at low temperatures has been studied experimentally in several other instances.[17-20] However, the *mfp* in those samples is shorter or the measurement temperature is higher than in Plombon, et al.'s samples and they do not clearly display the QMRS effects that we discuss.

$\rho_{QMRS}$ (Eq. 2) is temperature independent in a good metal like copper since $E_F \gg k_BT$. From our fit with $\delta w = 0.8$ nm, and with $\rho_{QMRS}$ estimated to be 0.28 μΩ cm for 73 nm lines at 300 K, the roughness contribution is 17 %. Assuming that $\delta w$ is independent of the nanoline width, its contribution would be one-half of the RK contribution for $w \sim 20$ nm

In summary, we consider the cause of energy dissipation in the important case of small linewidth copper lines used as interconnects in advanced electronic devices. Whereas previous efforts to understand the resistivity increase due to size effects only considered semiclassical effects, here we propose a quantum mechanical treatment. The model is used to analyze available electronic transport data for copper nanolines with widths in the range between 80 nm and 500 nm. The data is fit by assuming a fluctuation of the linewidth around its average value with a root-mean-square deviation of around 0.8 nm. Our model provides a framework for future studies of copper nanolines.

The work was supported by the Division of Materials Research of the U.S. National Science Foundation under Grant No. NSF-0611595 and NSF-0506842, by the Division of Materials of the U.S. Army Research Office under Grant No. DAAD4006-MS-SAH.

**FIGURE CAPTIONS**

Figure 1. (a) The data from Ref. 2 in a log-log plot. (b) The circles are respectively the low-temperature and room temperature resistivity of copper nanolines of various linewidths from Ref. 2. The dashed blue line is the fit according to the classical size effect in Eq. 1, including grain boundaries as presented in Ref. 2. The electron-surface scattering was assumed to be perfectly diffuse ($p = 0$), the grain size $g$ was taken as the linewidth, the boundary reflection coefficient $R = 0.25$ and the *mfp*s were 39 nm and 300 nm at room temperature and 20 K, respectively. The fits shown with a solid red line correspond to the case that the surface is only 90% diffuse, with $R = 0.18$, and considers the combined effect of semiclassical (finite size) effects (Eq. 1) and quantum mechanical roughness scattering (Eq. 2) with $\delta w = 0.8$ nm. These parameters are the same at 20 K and at 300 K. (c) RK specularity $p$ versus grain boundary reflection coefficient $R$ in Ref. 2 (blue) and in the present work (red).

Figure 2. Quantum Mechanical Roughness and copper nanolines. (a) Model. The line is divided into segments $i$ of length $L(i) \sim mfp$ and width $W(i)$ representing the average of the linewidth in the segment. Since the electron wavefunction is coherent over this region, surface roughness is considered quantum mechanically as an effective interaction potential that is derived from the boundary condition of the exact Hamiltonian. (b) The inset shows the band structure of copper nanolines. The solid and dashed horizontal lines represent levels of the subband structure of the metal in the line that are filled and unfilled, respectively. $\varepsilon$ is the spacing between energy levels in the nanoline that is $2\pi E_F / k_F w \sim 50$ meV. The gray band represents the thermal fluctuations at room temperature (300 K~30 meV). The dot-ended arrow represents a transition, of strength $\Delta$, caused by the effective surface roughness potential that leads to interband transitions within a band around the Fermi level. $\Delta$ can be estimated from $\Delta = \hbar \tau_{roughness} = \hbar(\rho_{bulk}/\rho_{roughness})\tau_{bulk}$ where $\tau_{bulk} = 1.9 \times 10^{-14}$ sec and therefore $\Delta \sim 100$ meV. The electron motion between segments is treated classically.

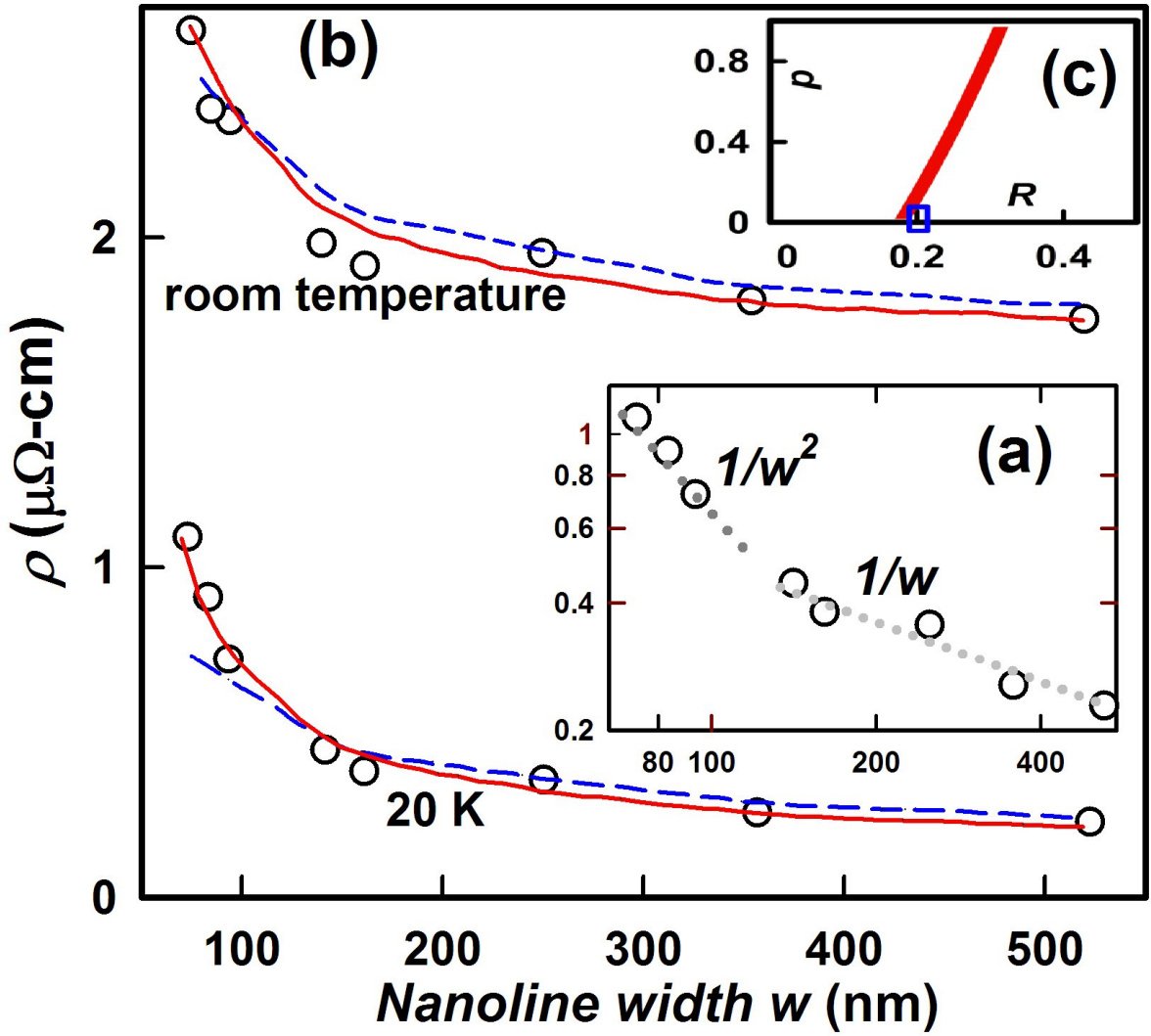

Huber. Cu nanolines. Figure 1. (2009)

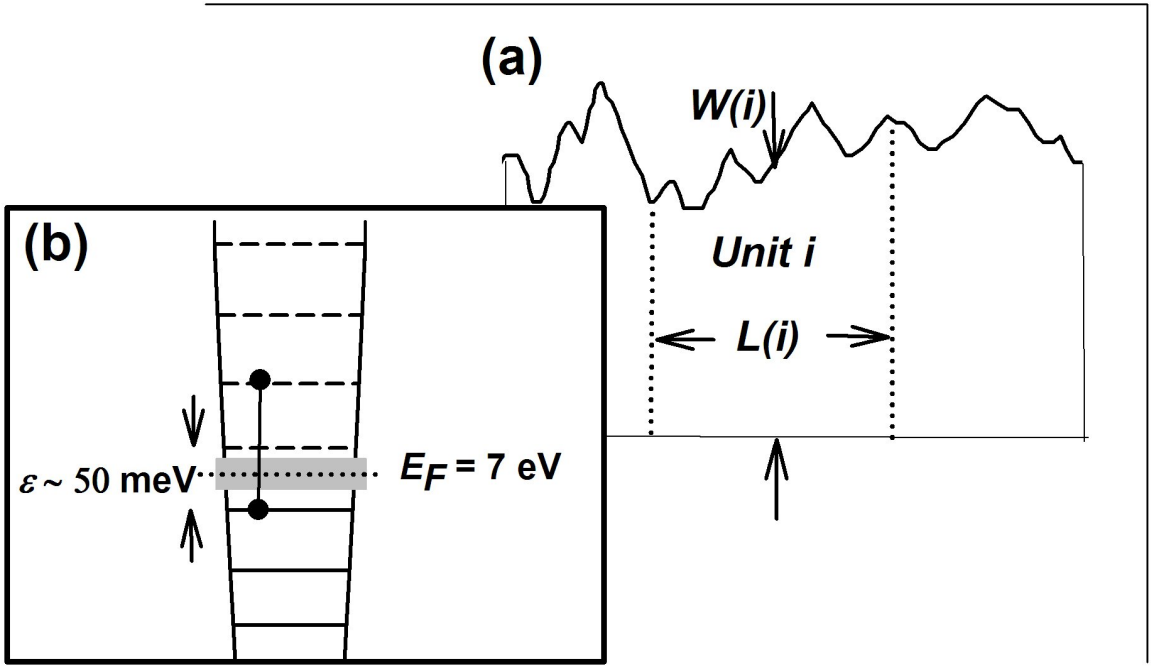

Huber. Cu nanolines Figure 2 (2009)